\documentclass[a4paper]{panl}

\usepackage{cite}
\usepackage{wrapfig}
\usepackage{graphicx}
\usepackage{amsfonts,amsmath,amssymb}
\usepackage{longtable}
\usepackage{rotating}
\usepackage{lscape}
\usepackage{epsfig}
\usepackage{multirow}
\usepackage{bm}
\usepackage[dvipsnames]{xcolor}
\usepackage{hyperref}
\hypersetup{%
colorlinks=true,%
citecolor=Blue,%
filecolor=Black,%
linkcolor=Blue,%
urlcolor=Blue %Violet
}
\usepackage[activate={true,nocompatibility},final,tracking=true,kerning=true,spacing=true,factor=1100,stretch=10,shrink=10]{microtype}
\microtypecontext{spacing=nonfrench}

\newcommand{\average}[1]{\left\langle{#1}\right\rangle}
\newcommand{\ceps}{\varepsilon}
\newcommand{\ud}     {\mathrm{d}}
\newcommand{\gev}    {\:\mathrm{GeV}}
\newcommand{\mev}    {\:\mathrm{MeV}}
\newcommand{\gevsq}  {\:\mathrm{GeV}^2}

\newcommand{\eq}[1]{Eq.(\ref{#1})}

\originalTeX
\begin{document}
% Journal sections (see http://pkp.jinr.ru/index.php/PEPAN_LETTERS/about/editorialPolicies#focusAndScope)
\issuearea{Physics of Elementary Particles and Atomic Nuclei. Theory}
% or in Russian
%\issuearea{ФИЗИКА ЭЛЕМЕНТАРНЫХ ЧАСТИЦ И АТОМНОГО ЯДРА. ТЕОРИЯ}

\title{%
Nuclear effects in the deuteron in the resonance and deep-inelastic scattering region\\%
Ядерные эффекты в дейтроне в резонансной и глубоконеупругой областях}
\maketitle
\authors{S.A.\,Kulagin$^{\:a,}$\footnote{E-mail: kulagin@ms2.inr.ac.ru}}
\from{$^{a}$\,Institute for Nuclear Research of the Russian Academy of Sciences\\%
60-letiya Oktyabrya Prosp. 7a, Moscow 117312, Russia}

\begin{abstract}
% Russian translation of the abstract
\noindent
Обсуждается гибридная модель неупругих структурных функций протона и нейтрона,
учитывающая как партонные так и резонансные структуры и применимая в широкой области значений масс $W$ рожденных адронных состояний. 
Для типичной кинематической области экспериментов JLab вычислены структурные функции дейтрона и показано хорошее согласие модельных предсказаний с экспериментальными данными.
Модель применяется для систематического исследования отношений $F_2^n/F_2^d$ и $(F_2^p+F_2^n)/F_2^d$
как в глубоконеупругой так и в резонансной областях в контексте анализа результатов эксперимента BoNuS в Jefferson Lab. \
\vspace{0.2cm}

%%%This is English translation of the abstract 
\noindent
We discuss a hybrid model of the proton and neutron inelastic structure functions
which incorporates both the partonic and the nucleon resonance structures and applicable in a wide region of the invariant mass $W$ of produced hadronic states.
Focusing at the typical kinematics of JLab experiments we compute the deuteron structure functions and demonstrate good performance of the model against data.
We perform systematic study of the ratios $F_2^n/F_2^d$ and $(F_2^p+F_2^n)/F_2^d$
for both, the deep-inelastic and the resonance, region in the context of recent measurements of BoNuS experiment at Jefferson Lab.
\end{abstract}
\vspace*{6pt}


%%%
%%%
%%%
\vspace*{-1ex}
\section{Introduction}
\label{sec:intro}

The BoNuS experiment at JLab recently reported a precise measurement 
of the ratio of the neutron and the deuteron structure functions
$F_2^n/F_2^d$ covering a wide region of Bjorken $x$
and 4-momentum transfer squared $Q^2$ between $0.7$ and $4.5\gevsq$~\cite{Tkachenko:2014byy}. 
Those data, in combination with the ``world'' data on the ratio $F_2^p/F_2^d$,
were used in Ref.\cite{Griffioen:2015hxa} in order to extract $R^d=F_2^d/(F_2^p+F_2^n)$
thus providing a measurement of the EMC effect in the deuteron.
The purpose of this paper is to compute the deuteron structure function
$F_2^d$ and study the ratios $R^d$ and $F_2^n/F_2^d$
in the kinematic region of BoNuS experiment. Note that BoNuS data were taken at relatively low values of
the invariant momentum transfer $Q$ and also a significant part of data fall into
the nucleon resonance region. For this reason in Sec.\ref{sec:pn} we develop a model
of the proton and the neutron
structure functions which incorporates both the partonic and the resonance structures
and spans a wide region of mass $W$ of produced hadronic states, from inelastic threshold to
the deep-inelastic scattering region.
In Sec.\ref{sec:d} we focus on the computation of the deuteron structure function $F_2^d$
and discuss model predictions for both the deep-inelastic
and the resonance regions. We then perform a detailed comparison of our predictions with
data of Refs.\cite{Tkachenko:2014byy,Griffioen:2015hxa}.

%%%
%%%
%%%
\vspace*{-2ex}
\section{The proton and the neutron structure functions}
\label{sec:pn}

For inelastic electron scattering off a proton (neutron) with four-momentum $p$ and four-momentum transfer $q$,
the different scattering regions are commonly characterized by the invariant mass squared of produced hadronic states
$W^2=(p+q)^2=p^2+Q^2(1/x-1)$, where $Q^2=-q^2$ and $x=Q^2/(2p\cdot q)$ is the dimensionless Bjorken variable.
The region of $W>2\gev$ and $Q>1\gev$ is referred to as deep-inelastic scattering (DIS),
in which the cross section is driven by scattering off quasi-free (anti)quarks in hadrons
described by the parton distribution functions (PDFs).
The structure functions (SF) depend on two independent variables, usually $x$ and $Q^2$.
A common framework to describe
DIS is the operator product expansion (OPE) which produce the power series in $Q^{-2}$ (twist expansion). 
In the first order, i.e. in the leading twist (LT), SFs are fully determined by PDFs.
The power corrections can be of two different types:
(i) contributions from higher-twist (HT) operators describing quark-gluon correlations
and 
(ii) correction arising from a finite nucleon mass (target mass correction, or TMC).
We also note that for the sake of computing the deuteron SF (see Sec.\ref{sec:d}), 
the nucleon SF are required in off-mass-shell region $p^2<M^2$, where $M$ is the nucleon mass.
Summarizing, the nucleon SF can be written as follows (which we will refer to as the DIS model)
\begin{equation}\label{eq:sfdis}
F_2^\text{DIS}(x,Q^2,p^2) = F_2^{\text{TMC}}(x,Q^2,p^2) + H_2(x)/Q^{2}, 
\end{equation}
where $F_2^{\text{TMC}}$ is the LT SF corrected for the target mass effects and $H_2$ describes the dynamical
twist-4 contribution (for brevity, we suppress explicit notation to the twists higher than 4).
In this paper the LT SF is computed using the proton and the neutron PDFs 
from a global PDF fit of Ref.\cite{Alekhin:2006zm,Alekhin:2007fh}.
We note that the studies of Ref.\cite{Alekhin:2006zm,Alekhin:2007fh}
were performed to the next-to-next-to-leading-order (NNLO) approximation
in QCD coupling constant with a special emphasis on a low-$Q$ region
and constrain the nucleon SF, including the HT terms, for $Q^2\geq 1\gevsq$.

In order to account for TMC, we follow Ref.\cite{Georgi:1976ve}.
Since the calculation of nuclear SF requires the nucleon SF in off-mass-shell region,
we analytically continue the equations of Ref.\cite{Georgi:1976ve} in the off-shell region
by replacing the nucleon mass squared $M^2$ with $p^2$. In particular, for $F_2$ we have
\begin{align}\label{eq:TMC}
F_2^{\text{TMC}}(x,Q^2,p^2) = &
\frac{x^2}{\xi^2 \gamma^3} F_2^\text{LT}(\xi,Q^2,p^2) +
\notag\\
& \frac{6x^3p^2}{Q^2\gamma^4}\int^1_\xi \frac{d\xi'}{\xi'^2}
\left[1+\frac{2xp^2}{\gamma Q^2}(\xi'-\xi)\right] F_2^\text{LT}(\xi',Q^2,p^2),
\end{align}
where $\xi=2x/(1+\gamma)$ is the Nachtmann variable  and $\gamma=(1+4x^2 p^2/Q^2)^{1/2}$.

It should be noted, that TMC procedure of Ref.\cite{Georgi:1976ve}
violates the $x\to1$ behavior leading to nonzero $F_2$ at and below the inelastic threshold 
(see, e.g., discussion in \cite{Kulagin:2004ie}).
The region of large Bjorken $x\sim 1$ corresponds to low $W$ and is affected by excitations of the nucleon resonances.
Clearly this is outside of the partonic picture of DIS.
In order to suppress contributions from low $W$ and formally obey the inelastic threshold requirement,
we multiply $F_2^\text{DIS}$ by the function $f_\text{th}(W)=1-\exp[(W_\text{th}-W)/d]$,
where $W_\text{th}=M+m_\pi$ with $m_\pi$ the pion mass.
We also assume that the parameter $d$ is of order of the pion mass and use $d=0.15\gev$ in the analysis discussed below.
This will ensure vanishing $F_2$ at the inelastic threshold $W=W_\text{th}$, and we also have
$f_\text{th}=1$ with a high accuracy for the $W$ values above the resonance region
(at practice this also holds for the second and the third resonance region).

In off-mass-shell region the structure function explicitly depends on the nucleon invariant mass squared $p^2$.
This dependence has two different sources:
(i) the terms $p^2/Q^2$ in \eq{eq:TMC} which lead to power terms at large values of $Q^2$
and 
(ii) nonpower terms from off-shell dependence of the LT SF.
Following Ref.\cite{Kulagin:1994fz,Kulagin:2004ie} we observe that for computing the nuclear SF
it would be enough to know the proton and the neutron SF in the vicinity of the mass shell $p^2=M^2$.
Then we use the nucleon virtuality $v=(p^2-M^2)/M^2$ as a small parameter and expand the SF in series in $v$.
To the leading order in $v$ we have
\begin{align}
\label{SF:OS}
F_2^\text{LT}(x,Q^2,p^2) &=
F_2^\text{LT}(x,Q^2)\left[ 1+\delta f(x,Q^2)\,v \right],
\\
\label{delf}
\delta f(x,Q^2) &= M^2\partial_{p^2}\ln F_2^\text{LT}(x,Q^2,p^2),  %\mid_{p^2=M^2},
\end{align}
where $F_2^\text{LT}$ on the right-hand side in \eq{SF:OS} is the structure
function of the on-mass-shell nucleon and $\partial_{p^2}$ in \eq{delf} denotes
the partial derivative with respect to $p^2$ taken on the mass shell $p^2=M^2$. 
The function $\delta f$ describes the relative
modification of the nucleon SF and PDFs in the vicinity of the mass shell.
A detailed study of nuclear DIS and Drell-Yan process in Refs.\cite{Kulagin:2004ie,Kulagin:2010gd,Kulagin:2014vsa}
and also recent global PDF analysis in Ref.\cite{Alekhin:2017fpf}
indicate no significant $Q^2$ as well as the nucleon isospin dependence of $\delta f$. 
Following these observations
we assume the function \eq{delf} to be $Q^2$ independent and  identical for the proton and the neutron, i.e.
$\delta f^{p,n}(x,Q^2)=\delta f(x)$.

The region $W<2\gev$ is driven by excitation of nucleon resonances which show up as pronounced structures
in the cross-sections. We first consider the proton SF. 
In order to model the proton SF in this region, we use the results of empirical fit 
of Ref.\cite{Christy:2007ve}, 
which takes into account the contribution from a few resonances as well as nonresonance background: 
\begin{equation}\label{eq:sfres}
F_2^\text{RES}(x,Q^2,p^2) = F_2^\text{CB}(Q^2,W^2).
\end{equation}
The SF computed by \eq{eq:sfres} will be refferred to as the RES model.
It should be noted that in the overlap region, in particular for $1.8<W<3\gev$ and $1<Q^2<9\gevsq$,
the results of the DIS~\cite{Alekhin:2006zm,Alekhin:2007fh} and the RES~\cite{Christy:2007ve}
fits are in a nice correspondence motivating us to use a combined DIS-RES model,
which spans a wide region of $W$. For a combined SF model we will use the DIS model (\ref{eq:sfdis}) for $W>W_2=2\gev$ and
the RES model (\ref{eq:sfres}) for $W<W_1=1.8\gev$. In order to insure the continuity of the resulting function,
we interpolate between the DIS and the RES models within the region $W_1<W<W_2$ using a linear in $W$ function
(the details of this model will be discussed elsewhere). 

It is worth mentioning that on average the DIS and the RES model give nearly equivalent description 
in the resonance region (the quark-hadron duality phenomenon~\cite{Bloom:1970xb}, for a review see Ref.\cite{Melnitchouk:2005zr}).
In particular, for the integral over the region from the inelastic threshold $W_\text{th}=M+m_\pi$ to
$W=W_2$ we find that the relation
\begin{equation}\label{eq:duality}
\int_{W_\text{th}^2}^{W_2^2} \ud W^2 F_2^{p(\text{DIS})}(x,Q^2) = 
\int_{W_\text{th}^2}^{W_2^2} \ud W^2 F_2^{p(\text{RES})}(x,Q^2)
\end{equation}
holds with a good accuracy in the region $1<Q^2<9\gevsq$ with a maximum difference between the DIS and the RES integral 
about 5\%.

Note that the discussion above refers to the proton and now we consider the model of the neutron SF. 
In the DIS region, the neutron LT SF is computed in terms of the proton PDFs relying on the isospin symmetry.
The isospin relations for the HT terms are not so obvious.
However, the isospin effect on the HT contribution was constrained phenomenologically from a global QCD analysis using 
proton and deuteron DIS data \cite{Alekhin:2003qq,Alekhin:2007fh,Alekhin:2017fpf} and we use these results in order to compute the neutron SF.
In the resonance region, an empirical model of the neutron SF was developed in Ref.\cite{Bosted:2007xd}.
In Ref.\cite{Bosted:2007xd} the neutron SF was obtained after subtraction the proton electroproduction data from corresponding deuterium data.
Comparing the results of Refs.\cite{Alekhin:2007fh} and \cite{Bosted:2007xd} in the overlap region,
we observe somewhat worse agreement for $F_2^n$, unlike a nice agreement for $F_2^p$.
A significant part of this disagreement arises from
a different treatment of smearing with momentum distribution and the binding effect in the deuteron 
in Refs.\cite{Alekhin:2007fh} and \cite{Bosted:2007xd}.  
In order to minimize this bias, we will model the neutron in the resonance region as follows.
We calculate $F_2^n$ for $W<2\gev$ using the RES fit of $F_2^p$ \cite{Christy:2007ve}
and the ratio $R_{np}=F_2^n/F_2^p$ computed using the DIS fit of Ref.\cite{Alekhin:2006zm,Alekhin:2007fh}. 
We aim to model the neutron SF down to the $\Delta(1232)$ region and inelastic threshold.
This region requires somewhat special consideration.
In particular, it follows from analysis of Ref.\cite{Bosted:2007xd} that $R_{np}\approx 1$ in the $\Delta(1232)$ region.
We therefore assume equal contribution to the proton and the neutron from the $\Delta(1232)$ resonance,
for which we use the notation $F_2^\Delta$, and consider
the following model for the neutron SF in the region $W<W_2$: 
\begin{equation}\label{eq:n-res}
F_2^{n(\text{RES})} = R_{np}\left(F_2^{p(\text{RES})} - F_2^\Delta\right) + F_2^\Delta,
\end{equation}
where the proton $F_2^{\text{RES}}$ and $F_2^\Delta$ computed using the fit of Ref.\cite{Christy:2007ve} 
while the ratio $R_{np}$ is computed using SF of the DIS fit of Ref.\cite{Alekhin:2007fh}.

%%%
%%%
%%%
\section{The deuteron structure function}
\label{sec:d}

In the region $x>0.15$ the inelastic scattering of leptons off nuclei is dominated by incoherent scattering off bound proton and neutron.
We consider the process in the target rest frame. The deuteron structure function $F_2^{d}$ can be written as follows
(for more detail see Ref.\cite{Alekhin:2003qq,Kulagin:2004ie}):
\begin{equation}
\label{eq:IA:2}
    F_2^{d}(x,Q^2) = \int \ud^3\bm p \left|\Psi_d(\bm p)\right|^2 K 
    \left[ F_2^p(x',Q^2,p^2)+F_2^n(x',Q^2,p^2) \right],
\end{equation}
where we consider $F_2^d$ as a function of $x=Q^2/(2Mq_0)$,
the integration is taken over the momentum of the bound nucleon $\bm p$
and $\Psi_d(\bm p)$ is the deuteron wave function in momentum space,
which is normalized as $\int\ud^3\bm p |\Psi_d(\bm p)|^2=1$.
Because of energy-momentum conservation, the four-momentum of the struck proton (neutron) is
$p=(M_d-\sqrt{M^2+\bm p^2}, \bm{p})$, where $M_d=2M+\ceps_d$ is the deuteron mass and 
$M=(M_p+M_n)/2$ the mass of the isoscalar nucleon and
$\ceps_d\approx -2.2\mev$ the deuteron binding energy.
We use the coordinate system in which the momentum transfer $\bm{q}$ is antiparallel to the $z$ axis,
$p_z$ and $\bm{p}_\perp$ are the longitudinal and transverse component of the nucleon momentum.
In this system the factor $K=(1+\gamma p_z/M)(1+{x'}^2(4p^2+6\bm{p}_\perp^2)/Q^2)/\gamma^2$,
where $p^2=p_0^2-\bm p^2$ and $x'=Q^2/(2p\cdot q)$ are the invariant mass
and the Bjorken variable of off-shell nucleon, respectively, 
and $\gamma^2=1+4x^2 M^2/Q^2$ \cite{Kulagin:2004ie}.

\begin{figure}[b!]
	\begin{center}
%%%
%%% Proton & Deuteron SF
%%%
%\vspace*{-2ex}%
%\hspace*{-1.5em}
\includegraphics[width=0.5\textwidth]{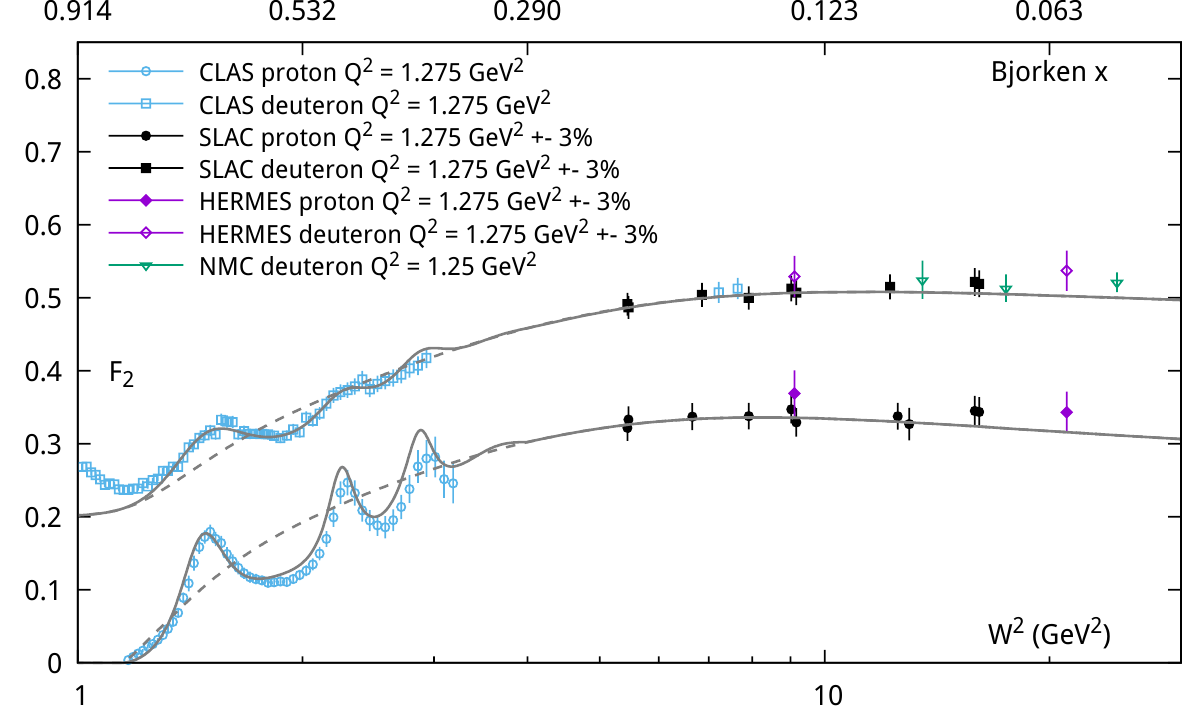}%
\includegraphics[width=0.5\textwidth]{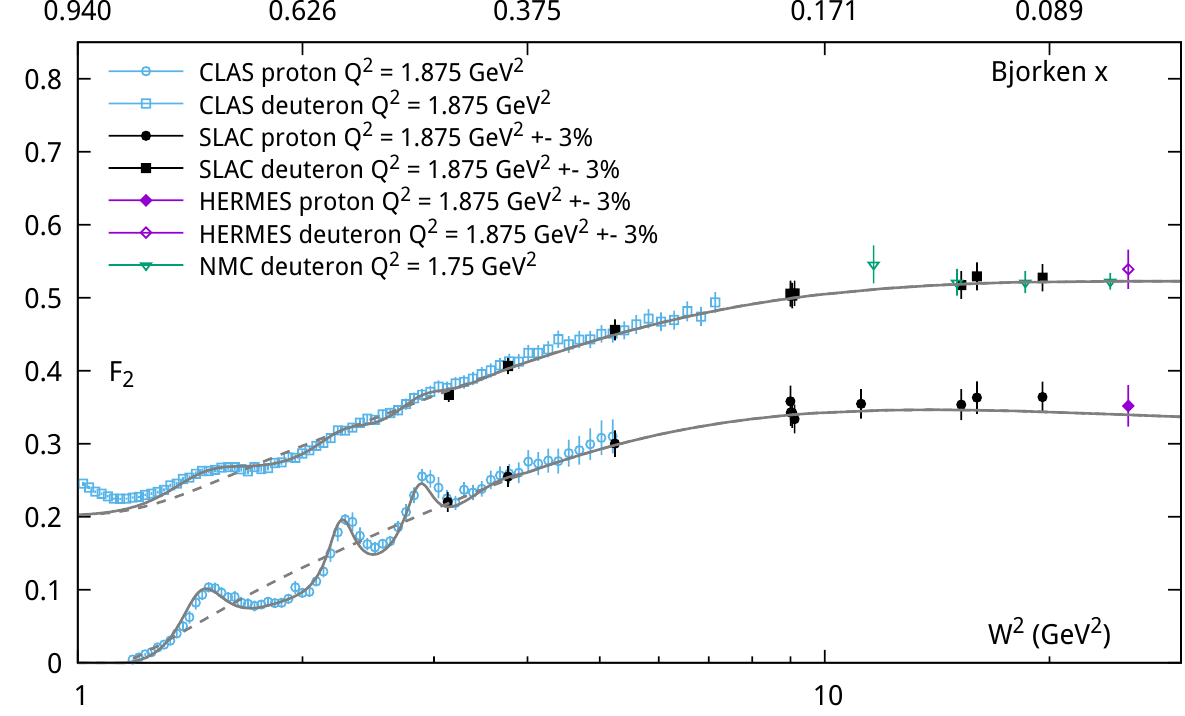}
%\hspace*{-1.5em}
\includegraphics[width=0.5\textwidth]{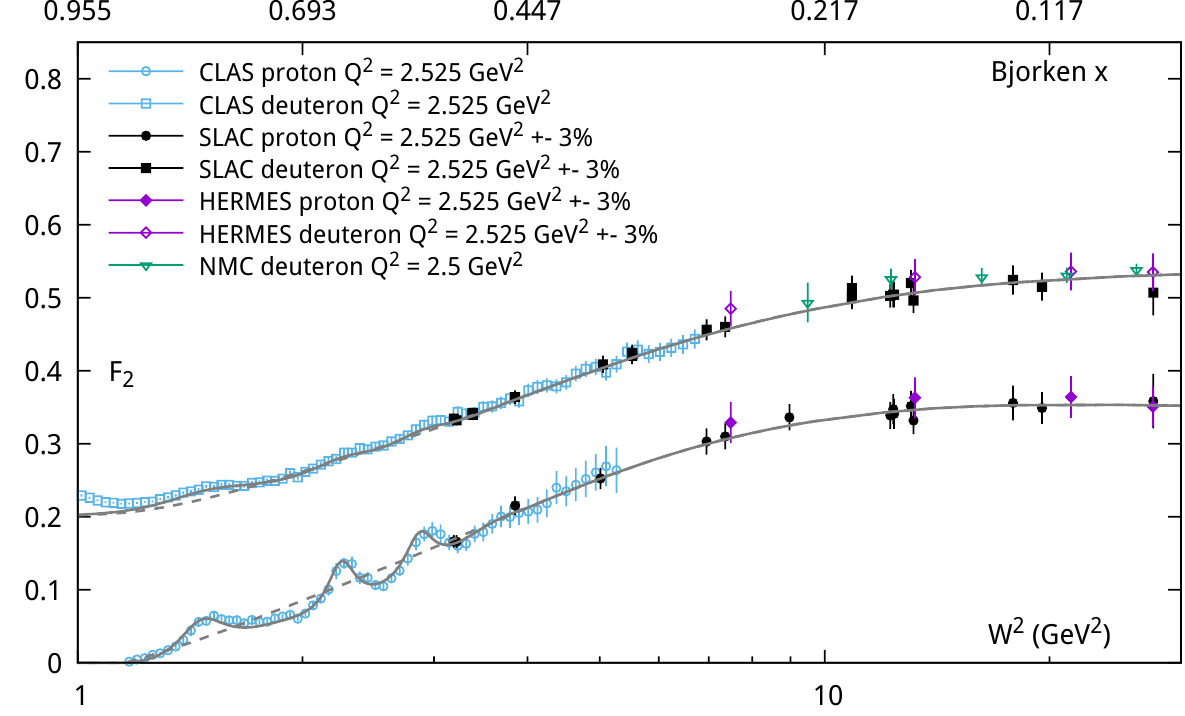}%
\includegraphics[width=0.5\textwidth]{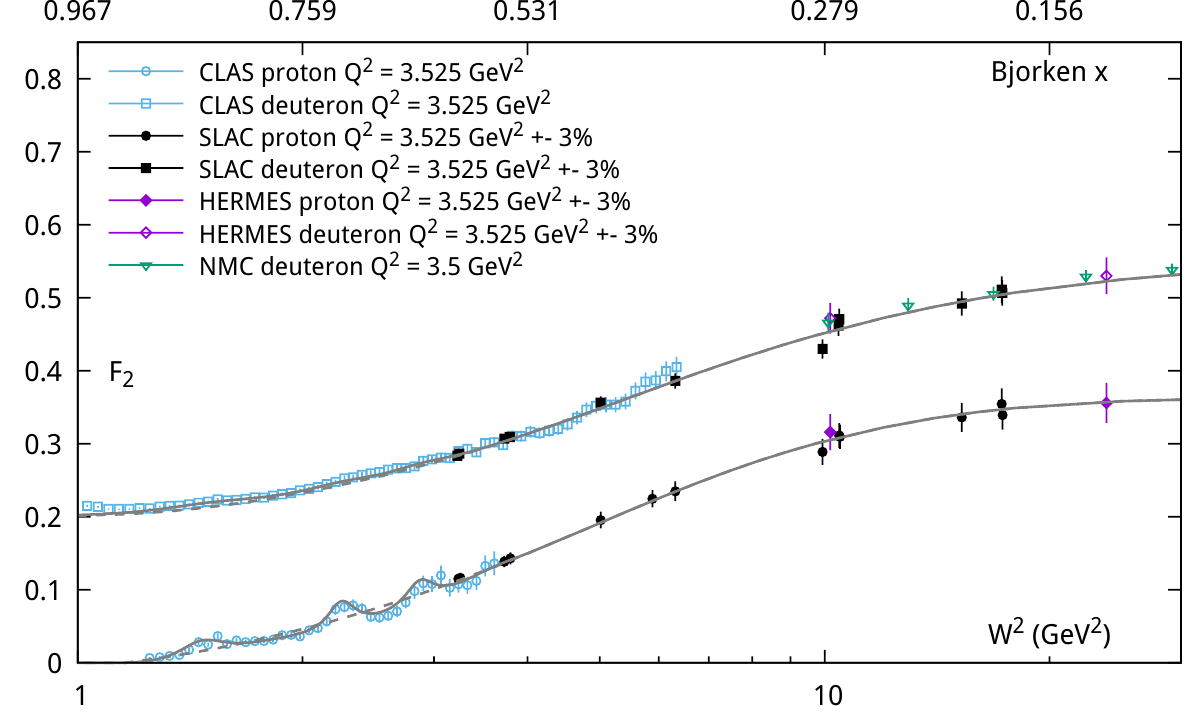}\vspace*{-1ex}
\caption{%
(Color online)
The proton and the deuteron structure functions as a function of $W^2$ computed at
$Q^2=1.275,\ 1.875,\ 2.525,\ 3.525\gevsq$ (the values of Bjorken $x$ are also shown in the upper $x$ axis).
The proton data points are from SLAC~\cite{Whitlow:1991uw}, JLab CLAS~\cite{Osipenko:2003bu}, and HERMES~\cite{Airapetian:2011nu} measurements,
while the deuteron data are from CLAS~\cite{Osipenko:2005gt}, NMC~\cite{Arneodo:1996qe}, and HERMES~\cite{Airapetian:2011nu}.
The data points are selected for the given value of $Q^2 \pm 3\%$.
For a better visibility the deuteron data are shifted up by 0.2.
The dashed and the solid curves respectively show the predictions of the DIS and the hybrid DIS-RES model discussed in the text.
\label{fig:f2pd}}
 	\end{center}
\vspace*{-4ex}%
\end{figure}

%%% Neutron over Deuteron

\begin{figure}[tb!]
	\begin{center}
%\vspace*{-2ex}%
%\hspace*{-1.5em}%
\includegraphics[width=0.5\textwidth]{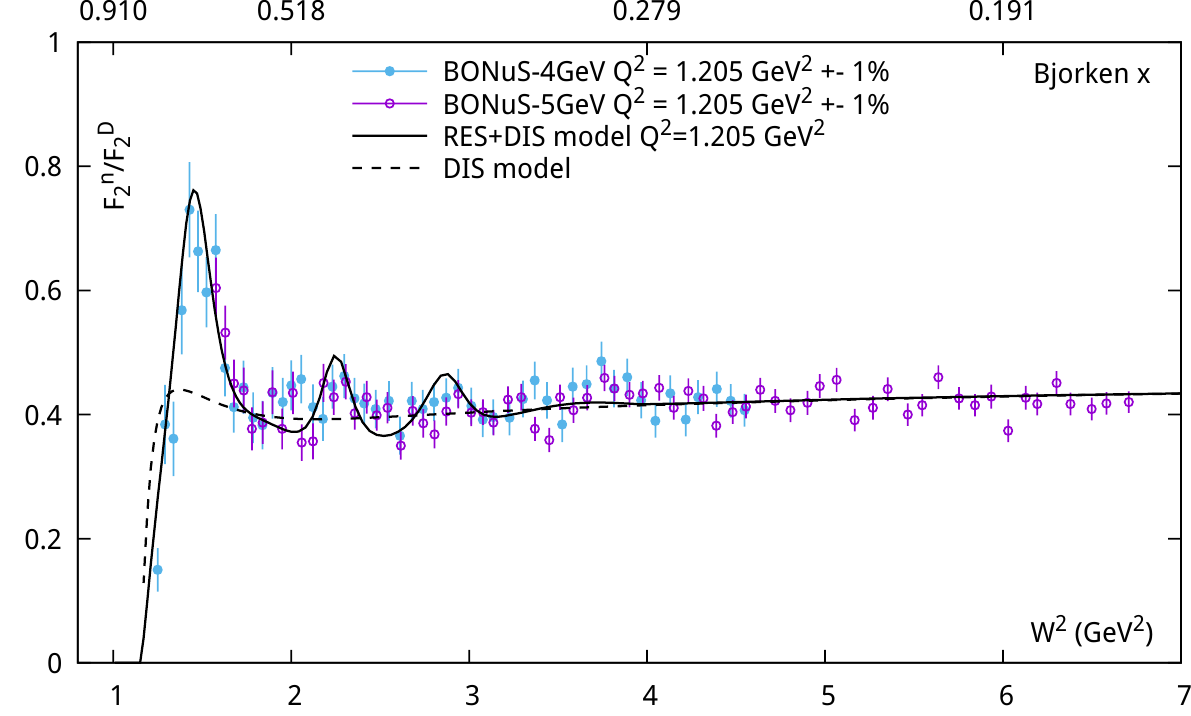}%
\includegraphics[width=0.5\textwidth]{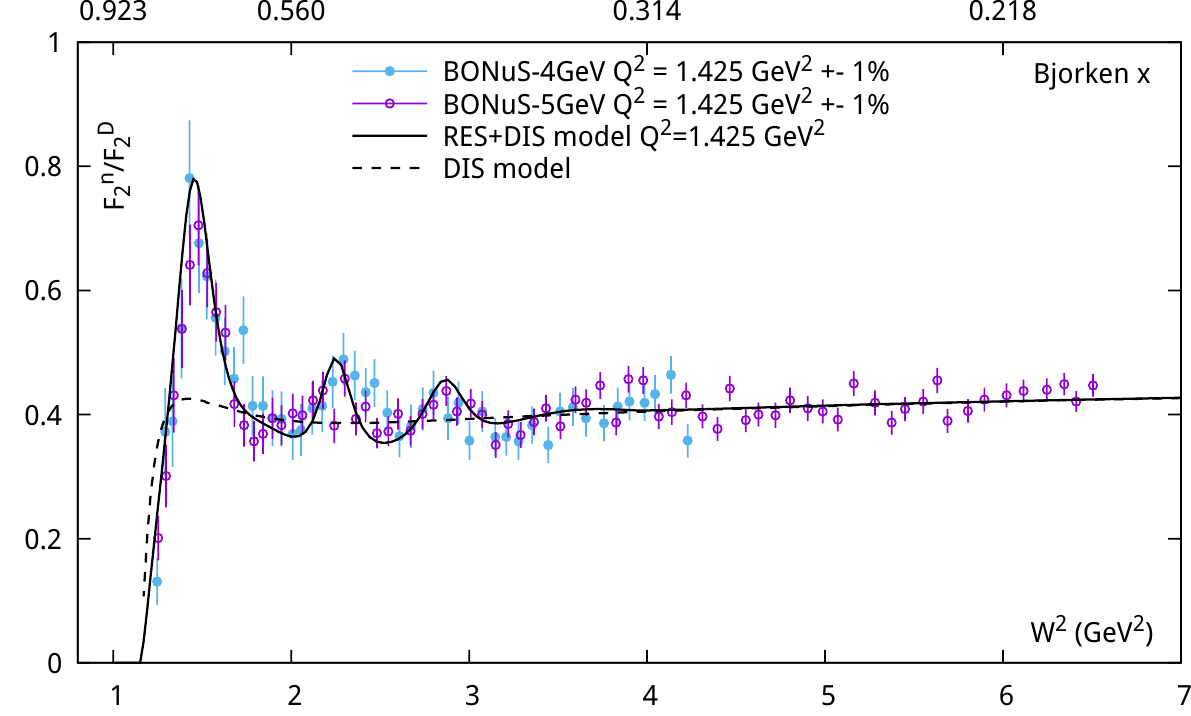}
\includegraphics[width=0.5\textwidth]{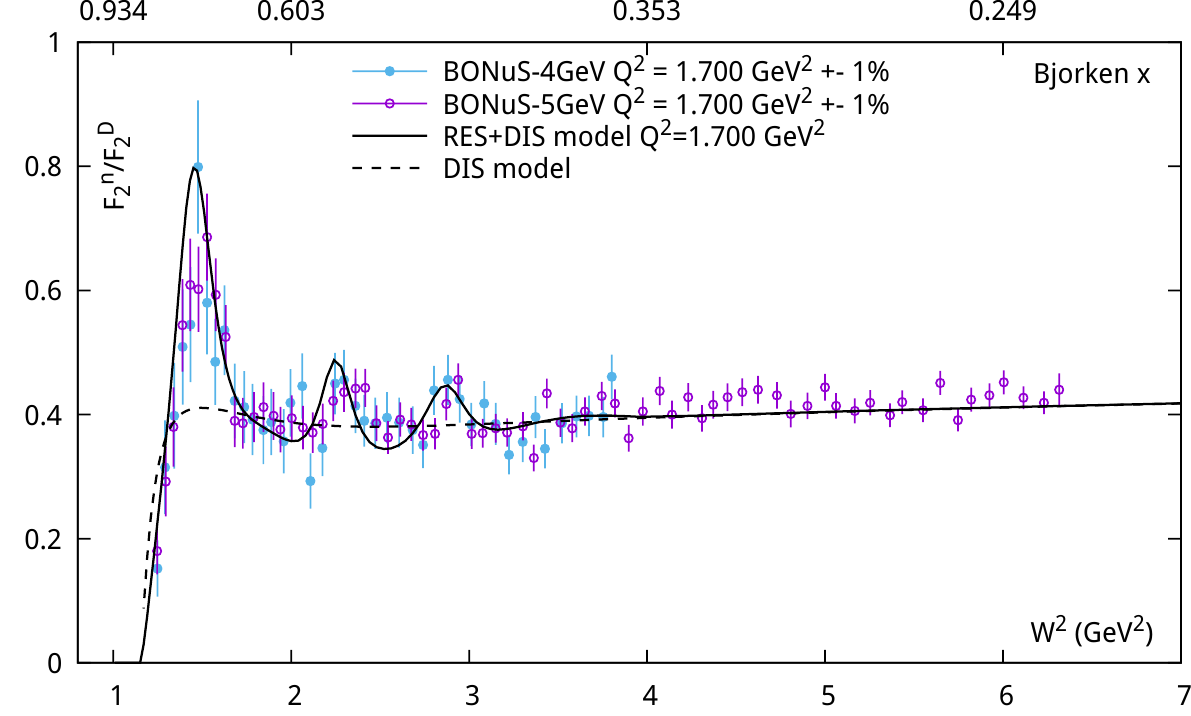}%
\includegraphics[width=0.5\textwidth]{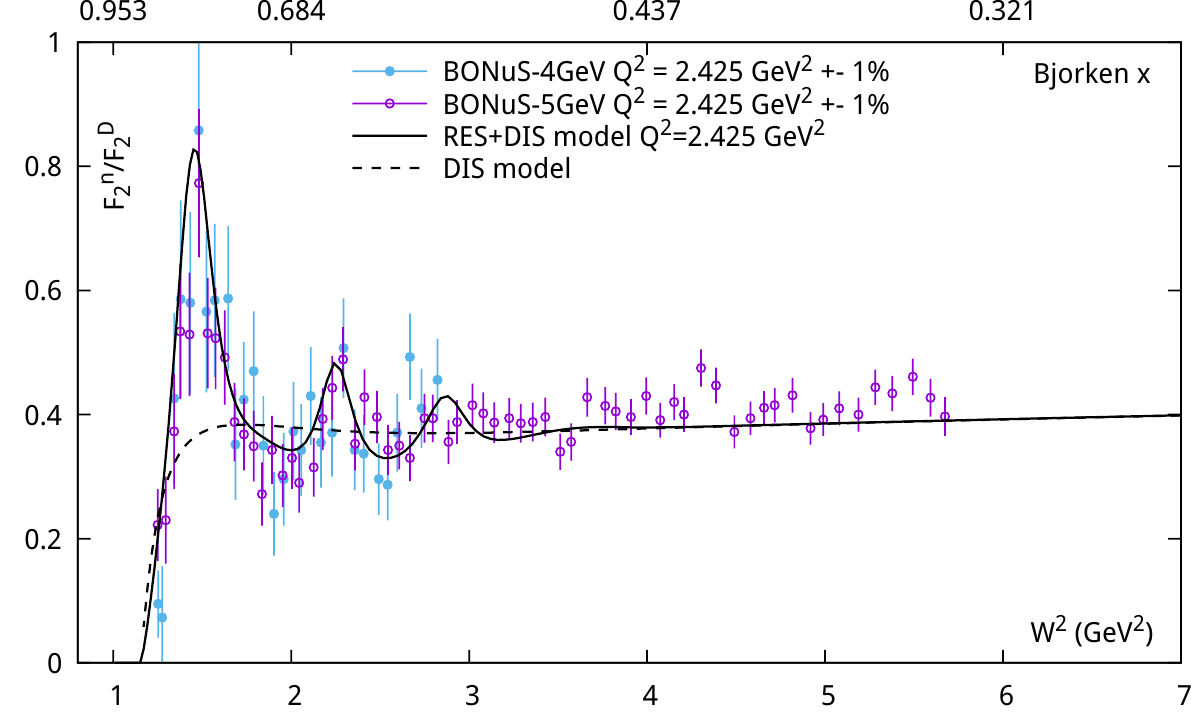}\vspace*{-1ex}
\caption{%
(Color online) 
The ratio of the neutron and the deuteron structure functions $F_2^n/F_2^D$ computed
at $Q^2=1.205,\ 1.425,\ 1.700,\ 2.425\gevsq$  as a function of $W^2$ (the values of Bjorken $x$ are also shown in the upper $x$ axis).
The data points are from the measurement by BoNuS experiment~\cite{Tkachenko:2014byy} at 4 GeV (closed circles) and 5 GeV (open circles)
selected for the given values of $Q^2\pm1\%$.
The dashed and the solid curves respectively show the predictions of the DIS and the hybrid DIS-RES model discussed in the text.
\label{fig:n_d}}
	\end{center}
\vspace*{-2ex}%
\end{figure}

We verify our approach by comparing the model predictions with various measurements.
Such a comparison is illustrated in Fig.\ref{fig:f2pd} to \ref{fig:n_d}.
In Fig.\ref{fig:f2pd} we show the predictions for the proton and the deuteron $F_2$ as a function of $W^2$
computed for a few fixed $Q^2$ together with the data from
SLAC~\cite{Whitlow:1991uw},
JLab~\cite{Osipenko:2003bu,Osipenko:2005gt},
NMC~\cite{Arneodo:1996qe}, 
and HERMES~\cite{Airapetian:2011nu} 
(the values of $Q^2$ in Fig.\ref{fig:f2pd} were selected such to maximize the overlap between data from different experiments).
We observe a good agreement between model predictions and data in a wide region of $W$ for both, the proton and the deuteron.
Figure \ref{fig:n_d} shows the results for the ratio $F_2^n/F_2^d$ of the neutron and the deuteron SF.
The model predictions are compared with data from BoNuS experiment~\cite{Tkachenko:2014byy} for a few fixed values of $Q^2$ shown in the plots.
From this comparison we observe a good agreement of our model with data in the full region of $W$. 
We also note a good performance of the model in the $\Delta(1232)$ resonance region indicating the validity of \eq{eq:n-res} for the neutron SF.%
\footnote{The details of the model and comparison with data will be discussed elsewhere.}

\begin{figure}[t]
%\vspace*{-8ex}%
	\begin{center}
%%%
%%%% pplusn/d in DIS and DIS-RES models at fixed Q2
%%%
\includegraphics[width=0.5\textwidth]{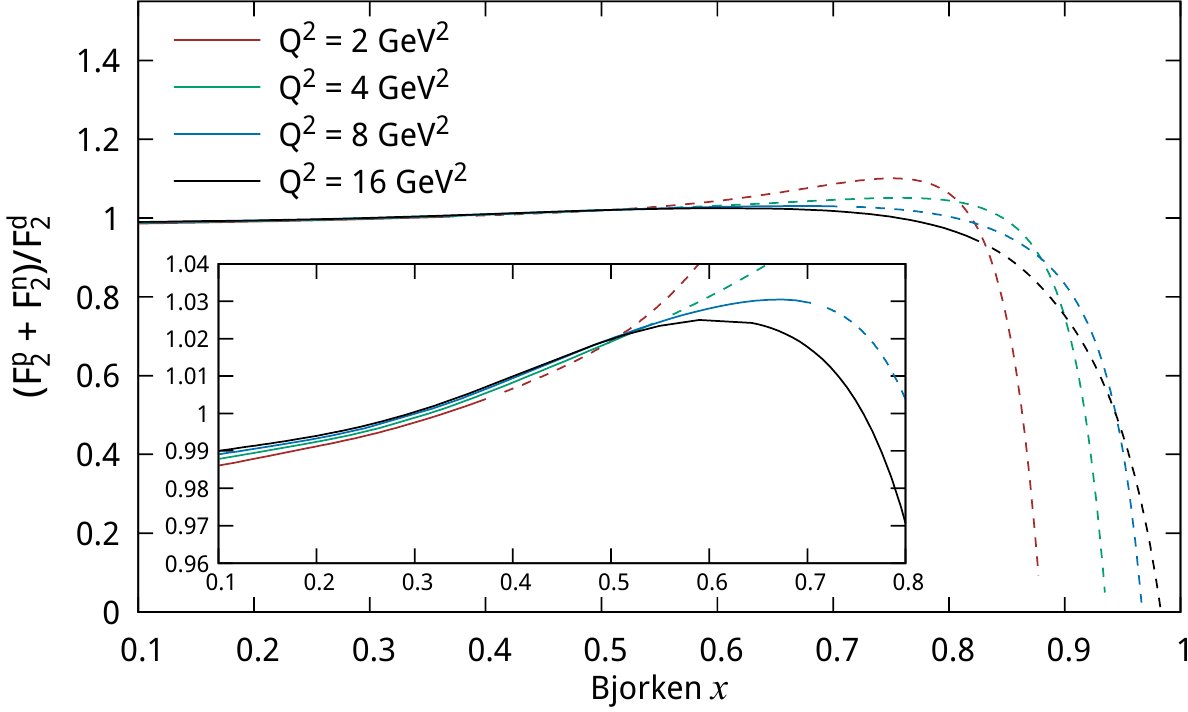}%
\includegraphics[width=0.5\textwidth]{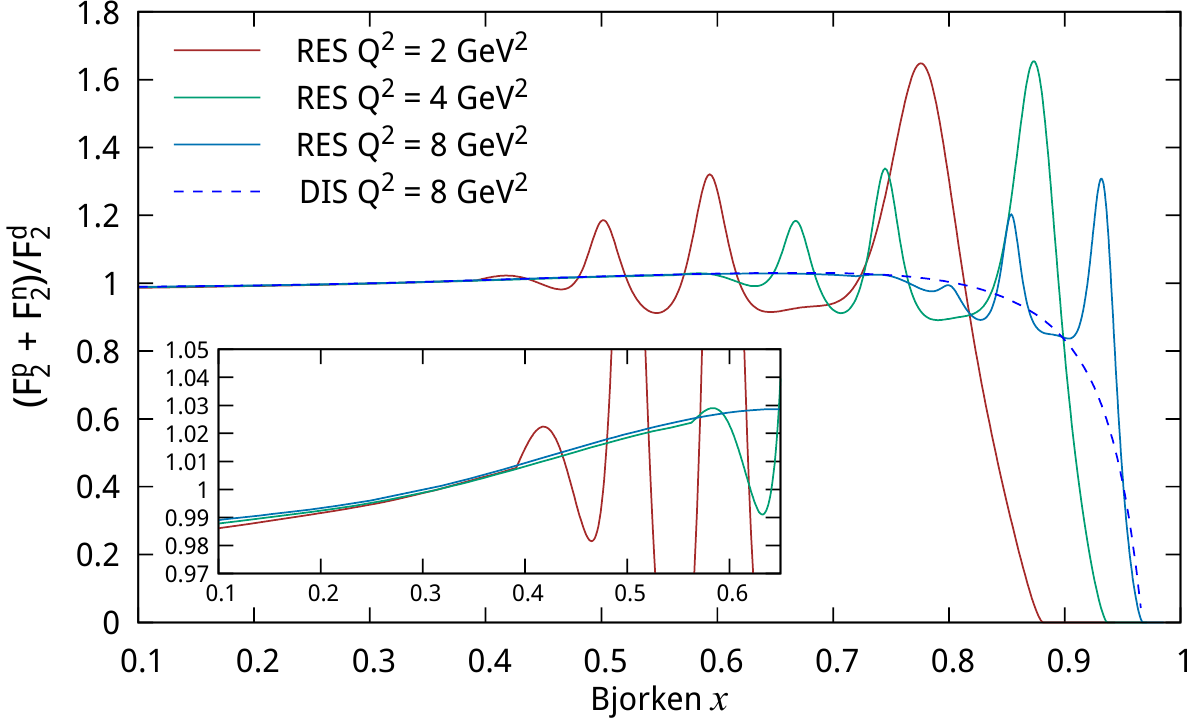}
\caption{%
(Color online)
The ratio $(F_2^p+F_2^n)/F_2^d$ computed as a function of Bjorken $x$
for $Q^2=2,\ 4,\ 8,\ 16\gevsq$ using the DIS model (left panel) and the combined DIS-RES model (right panel) discussed in the text.
For the DIS model, the solid lines indicate the region $W>2\gev$ while the dashed lines show the result 
from $W=2\gev$ down to inelastic threshold for each value of $Q^2$. 
The inset shows a magnified region $0.1<x<0.8$.\label{fig:rd}}
 	\end{center}
\vspace*{-2ex}%
\end{figure}

A good agreement with data allows us to proceed with a detailed study of the ratio $R^d$
which is traditionally used to measure the nuclear effects on the partonic level.
We focus on the region of relatively low values of $Q^2<10\gevsq$ and large Bjorken $x>0.1$,
which span the nucleon resonance region as well as the RES-DIS transition region.
The ratio $r_d=1/R^d$ computed using the DIS as well as a hybrid DIS-RES model  of Sec.\ref{sec:pn}
is shown in Fig.~\ref{fig:rd}.
In the left panel we show $r_d$ vs. Bjorken $x$ for the DIS model computed at a few different $Q^2$.
For $x<0.55$ we observe almost no $Q^2$ dependence, while the region of larger $x$
shows a strong $Q^2$ dependence which is because of the target mass correction. 
Note that $r_d$  has the inflection point at $x\approx 0.4$ at which $\partial_x^2 r_d=0$.
For this reason $r_d$ is almost a linear function of $x$ for $0.25<x<0.55$ with the slope $\partial_x r_d\approx 0.1$~\cite{Alekhin:2017fpf}. 
The latter is driven by the average energy $\ceps=p_0-M$ of the bound nucleon, $\average{\ceps}=\ceps_d-\average{T}$
with $\average{T}=\average{\bm p^2}/(2M)$ the average kinetic energy,
and by the average virtuality of the bound nucleon $\average{v}=2(\average{\ceps}-\average{T})/M$
\cite{Kulagin:1994fz,Kulagin:2004ie}.
We also note that the account of the nuclear binding correction 
\cite{Akulinichev:1985ij,Kulagin:1989mu}
together with off-shell effect allows us
to describe all available data on the nuclear EMC effect 
in heavy and light nuclei as discussed in detail in Refs.\cite{Kulagin:2004ie,Kulagin:2010gd,Kulagin:2016fzf}.

The results of our hybrid DIS-RES model are shown in the right panel of Fig.~\ref{fig:rd}.
For small and intermediate $x$ values, which correspond to $W>2\gev$, the
behavior of $r_d$ is identical to that of the DIS model.
For larger $x$ values, which correspond to the resonance region, $r_d$ shows pronounced oscillations
with the peaks' amplitude and position to be strongly dependent on $Q^2$.
For example, at $Q^2=2\gevsq$ the nuclear corrections can be as much as 60\% in the $\Delta$ resonance region
and reach about 30\% and 15\% in the second and third resonance region, respectively.
As $Q^2$ is rising the resonance curve approaches a smooth DIS curve, however
even at $Q^2=8\gevsq$ significant oscillations are present in the $\Delta(1232)$ region as well as in the second resonance region.

We consider $r_d$ instead of traditional $R^d$
in order to facilitate discussion at large values of Bjorken $x$,
as $r_d\to 0$ at the pion production threshold.
In this context we would like to emphasize that the behavior of $F_2^{p,n}$ near the inelastic threshold  has a strong impact on $r_d$
that allows us to test different models. 
In particular, TMC of Ref.\cite{Georgi:1976ve} violates the threshold behavior,
as was discussed in Sec.\ref{sec:pn}, and using \eq{eq:TMC} for the proton and neutron SF
in \eq{eq:IA:2} for low $Q^2 < 5 \gevsq$ 
would lead to unphysical values of $r_d>1$ at large $x$ near the inelastic threshold.
We recall that the factor $f_\text{th}(W)$ ensures $F_2^\text{DIS}$ to vanish on the inelastic threshold
and also has a significant impact on $r_d$ in the region of large Bjorken $x$. 
We also remark that the model of Ref.\cite{Christy:2007ve} respects the threshold behavior resulting in vanishing $F_2^\text{RES}$ at the inelastic threshold.

\begin{figure}[t!]
%\vspace*{-8ex}%
	\begin{center}
%%%
%%%% Comparison with BONUS data
%%%
\vspace*{2ex}%
\includegraphics[width=0.5\textwidth]{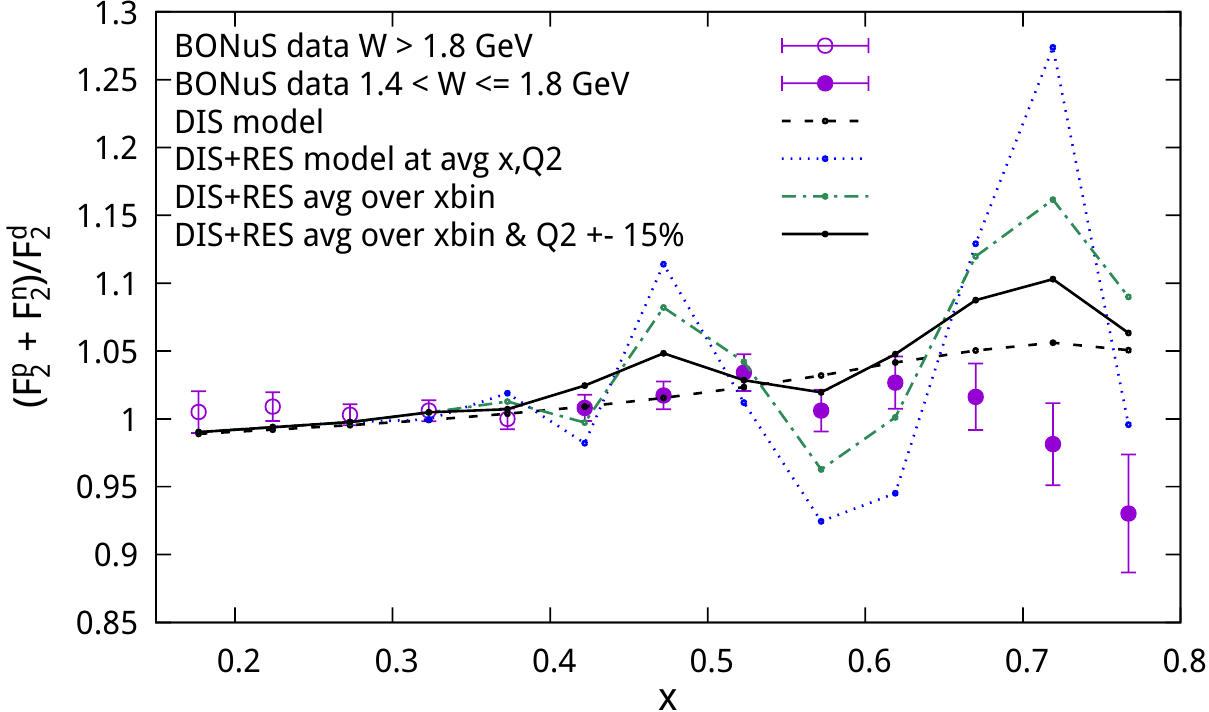}%
\includegraphics[width=0.5\textwidth]{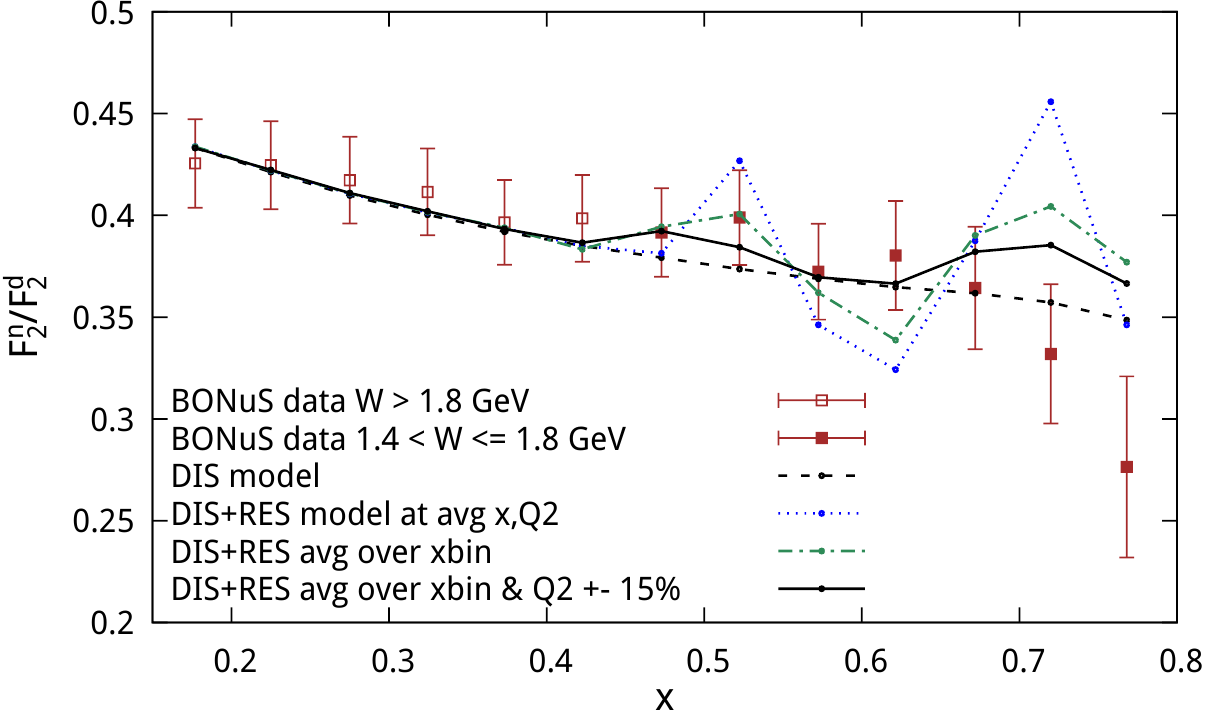}
\caption{%
(Color online)
Left panel: data points on the ratio $(F_2^p+F_2^n)/F_2^d$ from BoNuS experiment~\cite{Griffioen:2015hxa}
together with the result of analysis described in the text.
Right panel: data points on the ratio $F_2^n/F_2^d$ from from BoNuS experiment~\cite{Tkachenko:2014byy}
together with the result of analysis described in the text.
Solid symbols indicate the region of $W<1.8\gev$ while open symbols correspond to the DIS region $W>1.8\gev$.%
\label{fig:bonus}}
	\end{center}
%\vspace{-5mm}
\vspace*{-2ex}%
\end{figure}

We now compare in more detail the model predictions with recent data of BoNuS experiment on $r_d$ \cite{Griffioen:2015hxa} 
and $F_2^n/F_2^d$ \cite{Tkachenko:2014byy}.
The left panel of Fig.~\ref{fig:bonus} shows the data points from the measurement of Ref.\cite{Griffioen:2015hxa}.
The dashed line connects the points obtained with the DIS model using \eq{eq:IA:2} computed at the central values of
each $x$-bin with the corresponding average $Q^2$. 
The other lines show the predictions in our combined DIS-RES model using different assumptions about kinematics of the data points.
In particular, the dotted line is the result computed at the central values of each $x$-bin with the corresponding average $Q^2$,
similar to the DIS case (dashed line).
We observe the strong oscillation of $r_d$ for $x>0.4$ with rising amplitude which is due to the resonance behavior of the underlying
nucleon SF. These oscillations smooth out if we integrate (average) the structure functions over the $x$-bin (dashed-dotted line).
Further smoothing of the resonance behavior is observed if, in addition to the $x$-bin averaging, we average over $Q^2$.
The solid line is the result of such averaging assuming a uniform $Q^2$ distribution
for each of the data point around its central value within $\pm15\%$.
The right panel of Fig.~\ref{fig:bonus} shows the data points on the ratio $F_2^n/F_2^d$ 
as measured by the BoNuS experiment \cite{Tkachenko:2014byy} together with the result of similar analysis.

In summary, we discussed a hybrid DIS-RES model for the proton and neutron SF
basing ourselves on the results of two independent fits, respectively, in the DIS and the resonance region.
The model was applied to compute the deuteron SF
in the kinematic domain typical for JLab experiments with particular emphasis on the region $W<3\gev$.
We found good agreement with BoNuS data on the ratios $F_2^n/F_2^d$ and  $(F_2^p+F_2^n)/F_2^d$ for $W>1.8\gev$. 
For smaller $W$ these ratios are subject to strong oscillations due to resonance behavior of the nucleon SF.
The averaging over $x$ and $Q^2$ smooth out the resonance oscillations making the result close,
but not identical, to the prediction of the DIS model. 
A more detailed comparison with BoNuS data at very large $x$ would require further analysis.

I am grateful to S.Alekhin, V.Barinov and R.Petti for useful discussions 
and M.Osipenko for providing CLAS structure function data.
The work was partially supported by the grant of 
the Russian Science Foundation No.14-22-00161.

\bibliographystyle{pepan}
\bibliography{paper}

\end{document}